\documentclass[12pt]{iopart}

\usepackage{graphicx}

\usepackage{color}
\usepackage{amssymb}

\usepackage{dcolumn}						
\usepackage{bm}							
\usepackage[dvips]{epsfig}
\usepackage[dvips]{graphicx}					
\usepackage[T1]{fontenc}
\usepackage{amssymb}
\usepackage{latexsym}

\usepackage{ulem}

\usepackage{ifthen}

\ifthenelse{\boolean{false}}
{
\newcommand{\red}[1]{\textcolor{red}{#1}}    			
\newcommand{\blu}[1]{\textcolor{blue}{\it\tiny\bf [#1]}}    	
\newcommand{\rem}[1]{\red{\sout{#1}}}    			
\newcommand{\rep}[2]{\red{\sout{#1}}\red{#2}}
}{
\newcommand{\red}[1]{#1}    					
\newcommand{\blu}[1]{}    					
\newcommand{\rem}[1]{}    					
\newcommand{\rep}[2]{{}{#2}}
}

\begin{document}

\title[Experimental study of energy transport between two granular gas thermostats.]{Experimental study of energy transport between two granular gas thermostats.}

\author{Charles-\'Edouard Lecomte\rep{,}{ and} Antoine Naert}

\address{Laboratoire de Physique,  \'Ecole Normale Sup\'erieure de Lyon, Universit\'e de Lyon, C.N.R.S. UMR5672, 46 All\'ee d'Italie, 69364 Lyon Cedex 7, France.}
\ead{Antoine.Naert@ens-lyon.fr}

\begin{abstract} 
We report on the energy transport between two coupled probes in contact with granular thermostats at different temperatures. In our experiment, two identical blades, which are electromechanically coupled, are immersed in two granular gases maintained in different non-equilibrium stationary states, characterized by different temperatures. First, we show that the energy flux from one probe to another is, in temporal average, proportional to the temperature difference, as in the case of equilibrium thermostats. Second, we observe that the instantaneous flux is highly intermittent and that fluctuations exhibit an asymmetry which increases with the temperature difference. Interestingly, this asymmetry, related to irreversibility, is correctly accounted for by a relation strongly evoking the Fluctuation Theorem. As is, our experiment is a simple macroscopic realisation, suitable for the study of energy exchanges between systems in non-equilibrium steady states.
\end{abstract}

\section{Introduction}
The comparison between dissipative steady states and equilibrium states is a meaningful question in building the statistical physics of systems slowly relaxing toward equilibrium, or maintained in Non-Equilibrium Steady State (NESS) by external forces \cite{sasa2006, aumaitre2001}. Fluxes are essential in this situation, where steady states result often from the balance between boundary excitation and bulk dissipation \cite{aumaitre2001,falcon2008}. However, fluxes in this context often involve complex inhomogeneous and non-stationary transport processes \cite{aumaitre2006}. The theoretical treatment of irreversibility in transport is possible only in simplified situations \cite{derrida2007}. Although exactly solvable, these models involve elements which are very simple if compared to real phenomena. Besides, fluxes are very difficult to assess experimentally because conditions are often not properly constrained. In order to clarify the situation, we address experimentally the question of the flux between NESS in a very simplified configuration. Two independent thermostats are produced by vibrating at different amplitudes vessels filled with beads. They are coupled thanks to a simple electromechanical system. The resulting granular gases can be seen as two heat reservoirs at different temperatures, provided that the notion of temperature is properly extended to NESS. The temperature and flux are here accounted for by immersing in each vessel, a blade free to rotate around its vertical axis (the probe) and by measuring the fluctuations of its angular velocity  induced by the collisions with the grains.
The working definition adopted is \rem{the usual granular temperature, that has been founded as }\red{thus} an effective temperature, 
{resulting from the application of} the Fluctuation Theorem (FT) to the blade \cite{naert2012}, linked to the usual granular temperature. 
{If the systems, maintained at different temperatures $T_1$ and $T_2$, are coupled, energy is expected to flow from one to the other.}
\red{The situation is experimentally achieved by an electromechanical coupling between the blades.
Thus, the probes which are used to account for the temperature are also used to connect the heat reservoirs.}
{Doing so, we produce an }\rem{exact }analog of \rem{the experimental setup to study }heat conduction between equilibrium heat reservoirs at fixed temperatures. 
{However, crucial} differences are that the thermostats are dissipative\rem{,} and far from the thermodynamic limit, i.e. the number of particles and volume are not {\it very} large. \rep{In other words, f}{F}luctuations are 
{essential}, like in micro or nanoscale systems.\\
\indent
The \rem{experimental set-up and the }principle of the measurement\red{s and the experimental device} are described in
{~Sec.}~\ref{sec:principle and setup}. 
After a short discussion on the notion of temperature in Sec.~\ref{sec:energy flux}, 
{we report in Sec.~\ref{sec:temperature and energy flux}}
that the {temporal average of the} energy flux $\overline{\phi}$ is proportional to the temperature difference $\Delta T$.
The proportionality coefficient is derived explicitly from the system parameters. 
{Sec.~\ref{sec:fluctuations}} is devoted to the statistics of \red{the temporal fluctuations of }the \red{instantaneous }flux\red{, $\phi(t)$}.
\rep{The histograms}{Distributions} {of $\phi$} are highly non-Gaussian: their kurtosis is large, and their skewness increases with $\Delta T$,
\rep{accounting}{which accounts} for the increas\rep{ing}{e} of the mean. A striking observation is that the most probable value {of $\phi$} is always zero, whatever the average {$\overline{\phi}$}. 
In other words, these fluctuations are in general asymmetric, reflecting irreversibility of the energy flux in a temperature gradient. We consider in Sec.~\ref{sec:fluctuation theorem} this asymmetry of the fluctuations from the angle of the Fluctuation Theorem (FT). It is noted that the demonstration of the FT requires local time-reversibility, which is far from being valid in the present situation. The agreement with the FT, although not complete, goes far enough for such an original macroscopic system to excite curiosity.
These results are discussed in Sec.~\ref{sec:discussion}.
\section{Principle of the experiment and experimental setup}
\label{sec:principle and setup}
\begin{figure}[ht]
\includegraphics[width=7.5cm]{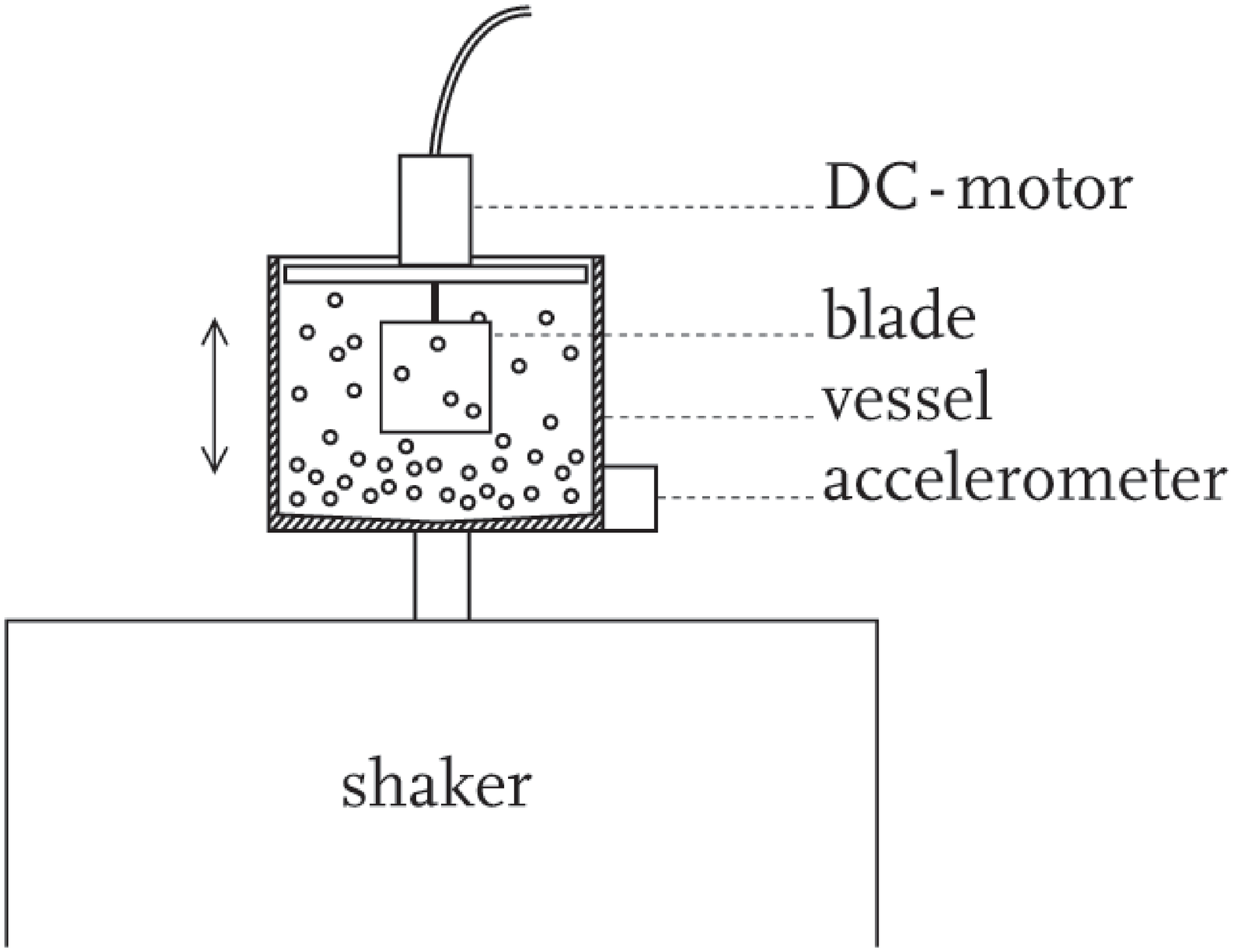}\hfill
\includegraphics[width=7cm]{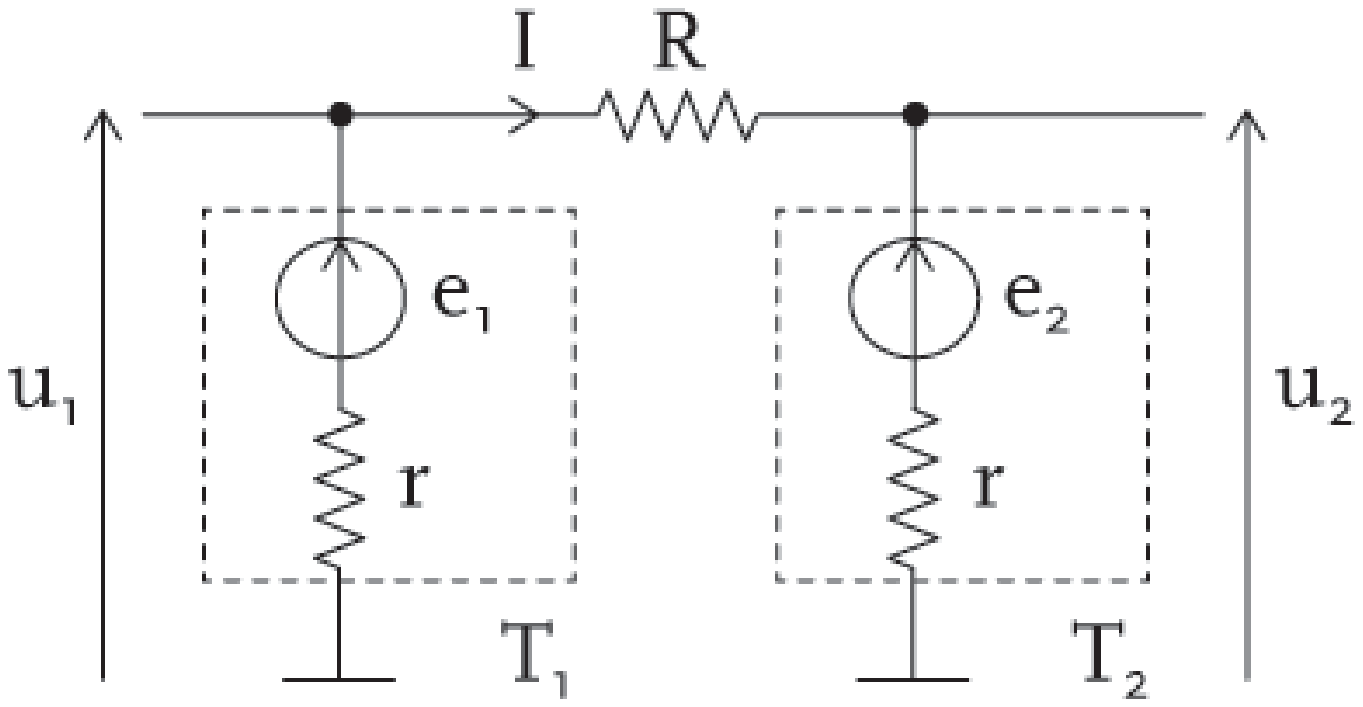}
\caption{On the left, one of the two ensembles shaker\,+\,vessel\,+\,motor\,+\,blade. The motor\,+\,blade ensemble is fixed on a cover. On the right is the electrical circuit of the compound system. The motors are schemati\rep{s}{z}ed by voltage sources ($e_i \propto \dot \theta_i$) and the internal resistances $r$. }
\label{fig:setup}
\end{figure}

\rep{The core principle of the experiment consists in the coupling of two identical copies of granular gas systems.}
{The core principle of the experiment consists in coupling two identical granular gases maintained at different granular temperatures and in measuring the resulting energy flux from one to the other.} \rep{Same number of beads are excited in two distinct vessels by identical shakers}
{Each of the granular gases consists of $N = 300$ stainless steal beads (diameter $3$~mm and mass $0.1$~g) }(\rep{see figure }{Fig.~}\ref{fig:setup}\red{,} left).
\rep{The vessels are}{The beads are placed in a} cylindrical \red{vessel }\rep{, $5$~cm diameter and $6$~cm height,}
{(diameter 5~cm and height 6~cm).}\rem{ with slightly conical bottom to improve mixing.}
\red{The gaseous phase is obtained by vibrating the vessel vertically
with a sinusoidal acceleration ranging from $2$ to $16$~g at 40~Hz.} (Shaker: Bruel~\&~Kjaer~4809.)
\red{In order to improve the mixing of the beads in the gas, we chose a vessel with a conical bottom.}
\rep{It is a duplication of the apparatus implemented in an earlier experiment}{Each of the granular gas systems is a duplication of that used in an earlier work} \cite{mounier2012}. \red{They are identical and only differ in the amplitude of the vibration, thus in their temperature.}
\rep{\noindent}{\newline\indent}
\rep{The coupling relies on an electromechanical symmetry, as for other studies by the same authors. The}
{In order to assess its temperature, a probe consisting of a thin}
blade\rep{s are dipped }{ is immersed} in{ each} granular gas (\rep{see figure }{Fig.~}\ref{fig:setup}\red{,} left).
\rep{Blades are made out of stainless steel}{The blade is a square}, $2$~cm $\times$ $2$~cm, \red{made out of a stainless steel plate (thickness $0.25$~mm).
It is attached to the vertical axis of a DC-motor, its lower edge positioned a few mm above the bottom of the vessel.}
Small brushed DC-motors of nominal power $0.75$~W \rep{were}{are} used (Maxon RE~10 118386).
The rotors are ironless to minimise inertia, and precious metal brushes improve the electrical contact with the commutator and reduce solid friction. 
A crucial issue is the symmetry of the two subsystems. They are constructed very carefully as identical as possible, from all points of view:  \\
-- Mechanically: dimensions are the same with tolerance always $100\,\mu$m, and smaller for the assembly of the blade on the axis of the motors to guarantee a tight mounting. All beads come from the same lot.\\
-- Electrically: all apparatus (generators, amplifiers, shakers) are the same (brands and models), and the recording is performed by two channels of the same acquisition board. The two DC-motors are the same, from the same lot. We keep them for the same operation time, for symmetric ageing. We note that the rotors do not rotate spontaneously like ratchets. As will be shown in the following, several quantities has been measured for different values of temperature difference. They all appear  symmetric for temperature difference inversion, which is equivalent to the permutation of the two subsystems. This symmetry is a strong indication that they are indeed identical.\\
\red{The temperature can be assessed from the fluctuations of the angular velocity $\dot \theta(t)$ of the blade.
\rep{T}{Indeed, t}he {\it granular temperature} is commonly defined, by simple analogy with the kinetic theory, as the variance of the particles velocity fluctuations \cite{goldhirsch2003}.
Using the same analogy, we write that the temperature of the blade, {\it thermalized} with the granular heat bath, 
is proportional to its average kinetic energy. Thus, for a system having a single degree of freedom, we write:
\begin{equation}
 \frac{1}{2}kT \equiv \frac{1}{2}M\overline{\dot{\theta}^2}
 \label{eq:temperature}
\end{equation}
where $M$ stands for the total moment of inertia of the parts in solid rotation (blade + rotor of the motor, $M = 3.33\,10^{-8}$~kg m$^2$).
Eq.~(\ref{eq:temperature}) defines the energy scale $kT$ which accounts for the {\it temperature} of the system, here the blade in contact
with a stationary granular heat bath. Note that this 'temperature' or 'thermal energy' will be used indeterminately for $kT$ in the following. We choose not to express temperature in Kelvins, because the Boltzmann constant $k_{\rm{B}}$ is meaningless in such macroscopic system.}\rep{\\}{\newline\indent}
The granular gas is very dilute, in the sense that the mean free path is not much less than the size of the vessel of radius $R$, or even the blade's dimension. It is difficult to investigate experimentally the bulk of granular matter. But, assuming in first approximation and for small acceleration of the vessel that most of the beads are in the lower $h\simeq 1$~cm because of stratification, an average distance between particles is calculated from the average density $n=N/(\pi R^2h)$, of the order of $4$~mm. Thanks to kinetic theory, the mean free path is: $\lambda=1/(n\pi r^2)$, where $r = 1.5$~mm is the radius of a bead. This gives a rough lower estimate $\lambda\sim 9$~mm, as discussed already in \cite{mounier2012}. This value depends very much on the acceleration given to the vessel, i.e. temperature, but confirms the assertion that the granular is dilute in any case of interest here.\\
\indent The coupling relies on an electromechanical symmetry. In practice, the two subsystems are coupled electrically by \red{connecting the motors with one another by}
a resistor $R=23.3~\Omega$, approximately equal to the internal resistance $r$ of the motors (\rep{see figure }{Fig.~}\ref{fig:setup}\red{,} right).
The \red{internal }resistance of the motors\red{, $r\simeq22~\Omega$,} \rep{are}{is} simply measured with an ohmmeter\red{, at room temperature} 
in the absence of granular gas. (We checked that, in what follows, the inductance of the motors can be neglected.)
The principle of the coupling is the following. A DC-motor can be used reversibly as motor or generator. As a motor, the torque\rep{ $\Gamma$}{, $\Gamma=\alpha\,I$,} is proportional to the current\red{, $I$}\rem{ : $\Gamma=\alpha I$.}
(The factor $\alpha$ represents all the physical characteristics of the motor, like number of poles, coils, and turns, magnetic field of the permanent magnets, and geometric factors\rem{.})\red{.}
As a generator, a voltage\red{, $e=\alpha\,\dot\theta$,} proportional to the angular velocity\red{, $\dot \theta$,}  is induced\rem{: $e=\alpha \dot \theta$.}
\red{(Notice here that the coefficient $\alpha=4.27\,10^{-3}$~Vs/rad is unique).}
\rem{Since the blade is free to rotate, an energy $\dot w_i$ can be exchanged between the motor $i$ and the surrounding bath at $T_i$, in one way or the other ($i$ numbers the subsystem). Each time unit, the work performed is: $\textcolor{pink}{\dot w_i}=\pm\Gamma_i \dot \theta_i=\pm I e_i$. By convention the power is counted positive when it flows from the bath to the motor. Note that no calibration is needed, as the prefactor $\alpha$ is unique. }
\rep{W}{Thus, w}hen momentum is \rep{given}{transferred} to one blade by the surrounding beads\red{ in one reservoir}, a voltage is induced in the circuit. This voltage causes a current to circulate in the other motor. A torque is \red{then }produced on the second blade, that transfers momentum to the other reservoir. As the electromechanical devices are reversible, this process occurs randomly in both \rep{ways}{directions}\rem{, but a net energy flux appears from the gas of higher to lower temperature}.\red{\\}
\red{\indent Assessment of the temperatures and of the energy flux rely on the measurement of electrical quantities, only.}
\rep{Everything is measured from}{The} voltages $u_1$ and $u_2$ \rep{directly:}{are monitored (NI PXI-4462) at $1$~kHz for 1 hour samples.} 
\red{Considering the electrical circuit in Fig.~\ref{fig:setup}, we write:}
\begin{equation}
\eqalign{
e_1&=u_1+rI,\cr
e_2&=u_2-rI,}
\label{eq:ddp} 
\end{equation}
\rep{with}{and:} 
\begin{equation}
I = \frac{1}{R_{\rm{tot}}}\left(e_1-e_2\right) = \frac{1}{R}\left(u_1-u_2\right)
\label{eq:current} 
\end{equation}
where $R_{\rm{tot}}$ is the sum of all resistances in the circuit: $R_{\rm{tot}}=2r+R$.
Notice that the knowledge of $u_1$ and $u_2$ is enough to assess, for instance, the temperature $kT_1$.
Indeed, from Eq.~(\ref{eq:temperature}), $kT_1 = M\overline{\dot{\theta_1}^2} = \frac{M}{\alpha^2}\overline{{e_1}^2}$
with, from Eqs.~(\ref{eq:ddp}) and (\ref{eq:current}), $e_1 = u_1 + \frac{r}{R}\,(u_1 - u_2)$. A similar relation exists
for $kT_2$. Thus, the temperature difference can be obtained from $u_1$ and $u_2$ by using:
\begin{equation}
kT_1 - kT_2 = \frac{M}{\alpha^2}\,\frac{R_\mathrm{tot}}{R}\,\Bigl(\overline{{u_1}^2}-\overline{{u_2}^2}\Bigr)
\label{eq:temp_diff}
\end{equation}
\red{In Sec.~\ref{sec:temperature and energy flux}, we report on the dependence of the energy flux
in the electrical circuit as consequence of such temperature difference.}

\red{\section{Energy flux}
\label{sec:energy flux}}

\red{Let us first discuss thoroughly the quantities we shall measure.}
\rep{A}{Notice that, a}lthough \rem{some }energy is dissipated \rep{during the transfer}{in the resistors},
the current is conserved \rep{along the loop}{in the circuit that insures the coupling}.
This is a decisive difference with \red{classical} heat conduction \rep{system}{between thermostats} in presence of losses.
As a consequence of this conservation, the \red{magnetic} torques $\Gamma_1$ and $\Gamma_2$ resulting from the current $I$
applied respectively to the blade 1 and 2 are equal at all times ($\Gamma_1 = \Gamma_2 = \Gamma$).
Thus, 
the instantaneous electromagnetic power on blade 1 can be written $\dot w_{1} = \Gamma \dot \theta_1 = e_1 I$,
whereas, for blade 2, $\dot w_{2} = \Gamma \dot \theta_2 = - e_2 I$ (The change in sign is only due to the orientation of the current $I$). 
Therefore, the current $I$, even if it imposes the same torque $\Gamma$ to both blades,
is associated to a difference in the electromagnetic power transferred to the blades,
$\Delta \dot w \equiv \dot w_{1} - \dot w_{2}$, which is only due to the difference $(\dot \theta_1 - \dot \theta_2)$
in their instantaneous velocity of rotation.

Considering now that the current $I$ originates itself from the fluctuations of the angular positions of the blades,
one can regard the difference $-\Delta \dot w$ as the difference between the power transmitted by the blade 1 to the blade 2
and the power transmitted by the blade 2 to the blade 1,
thus as the energy flux, $\phi$, between them. In electrical variables, we have: 
\begin{equation}
\phi = - \Delta \dot w = I\,(e_1+e_2) = \frac{1}{R_{\rm{tot}}}\left(e_1^2-e_2^2\right).
\label{eq:flux} 
\end{equation}
\section{Temperature and mean energy flux}
\label{sec:temperature and energy flux}
\red{\subsection{Temporal average}}
\label{sec:average}
\rem{The {\it granular temperature} is commonly defined as the variance of the particles velocity fluctuations: $T\equiv \overline{v^2}$, by simple analogy with the kinetic theory \cite{goldhirsch2003}. However, its founding is weak, as no equipartition hypothesis can be invoked out-of-equilibrium. The average kinetic energy of a 1-degree of freedom object {\it thermalised} with the granular heat bath is proportional to this temperature: $\frac{1}{2}kT=\frac{1}{2}M\overline{\dot{\theta}^2}$. The following working definition has been adopted: $k\,T_i=\frac{M}{\alpha^2}\,\overline{e_i^2}$, where $\alpha$ is the proportionality coefficient between induced voltage and velocity ($\alpha=4.27\,10^{-3}\,\rm{V/rad}\,\rm{s}^{-1}$) and $M$ is the total moment of inertia of the blade ($M_{b}=2.67\,10^{-8}\,\rm{kg}\,\rm{m}^2$) and the rotor ($M_{r}=0.66\,10^{-8}\,\rm{kg}\,\rm{m}^2$). }
Recalling \rem{equation \ref{eq:temperature}}{the definition of the temperature Eq.~(\ref{eq:temp_diff}), we write the temporal average of Eq.~(\ref{eq:flux}) in the form:}
\begin{equation}
\eqalign{
\overline{\phi}
&=\frac{\alpha^2}{M R_{\rm{tot}}}\left(kT_1-kT_2\right).
}
\label{eq:fourier}
\end{equation}
\rem{The coefficient equivalent to the heat conductivity is only valid in this experimental configuration. }
The energy flux is proportional to the temperature difference. The transport coefficient only depends
on the characteristics, $\alpha$, $M$ and $R_\mathrm{tot}$ of the coupling.
The result is exact, provided that $kT_1$ and $kT_2$ are measured in place on the coupled systems,
or that the systems are not altered by the coupling, i.e. that the blades remain in equilibrium with
unaltered heat baths. This assumption is reasonable as $\overline{\phi}$ is extremely small (of the order of $10^{-5}$~W)
if compared to the injected power (of order $10$~W) needed to keep the granular gases in motion i.e. compensate its dissipation.
We checked experimentally that the difference in e.m.f. with free fluctuations, assessed independently by measuring the
voltages $e_1$ and $e_2$ in open circuit, or in the system of coupled motors is indiscernible. 
The prefactor ${\alpha^2}/({M R_{\rm{tot}}}) = 8.13$~Hz is known precisely. The parameters $\alpha$ and the moment of inertia of the motor are read in Maxon's data sheet, and the moment of inertia of the blade is calculated on the basis of precise dimension and weight measurements. 
The measurement of the set \{$u_1(t), u_2(t)$\}  give on one hand $I(t)$ provided $R$, and in the other hand \{$e_1(t), e_2(t)$\} provided internal resistance of the motors $r$. $R$ is easily and precisely obtained, so the quality of the measurement lean on the reliability of the measurement of $r$. Being among the main sources of errors, it concentrates our attention and the difficulty of this measurement, otherwise rather easy, as long as large but homogeneous time series can be acquired. The values $r$ are measured before and after each time series recording such as to account for temperature elevation in the actual excitation condition, after a few hours of operation.) Another source of error inherent to DC-motors is friction losses. They have been evaluated to be about an order of magnitude less than a typical power $\overline{e_i I}$.\\\indent
\red{It is remarkable that a non-zero average energy flux circulates from one bath to the other, although torques, $\Gamma_i$, as well as angular velocities,
$\dot\theta_i$, are always $0$ in average.
This non-intuitive feature results from the fact that correlations exist between the voltages $e_i$ and the current $I$ (and thus between $\Gamma$ and $\dot\theta_i$),
caused by the conservation of the current $I$ over the whole circuit.}
\red{The resulting average energy flux between the blades in contact with granular thermostats is proportional to the temperature difference}.
\red{This is similar to what would be observed for the transport between equilibrium thermostats.}
\rep{Let us consider i}{However, as shown i}n the following\red{,} the fluctuations of the instantaneous energy flux\red{,} $\phi(t)$, \rep{which are also accessible experimentally}
exhibit significant differences with equilibrium systems.   

\subsection{Temporal fluctuations}
\label{sec:fluctuations}
\begin{figure}[h!]
    \includegraphics[width=7.8cm]{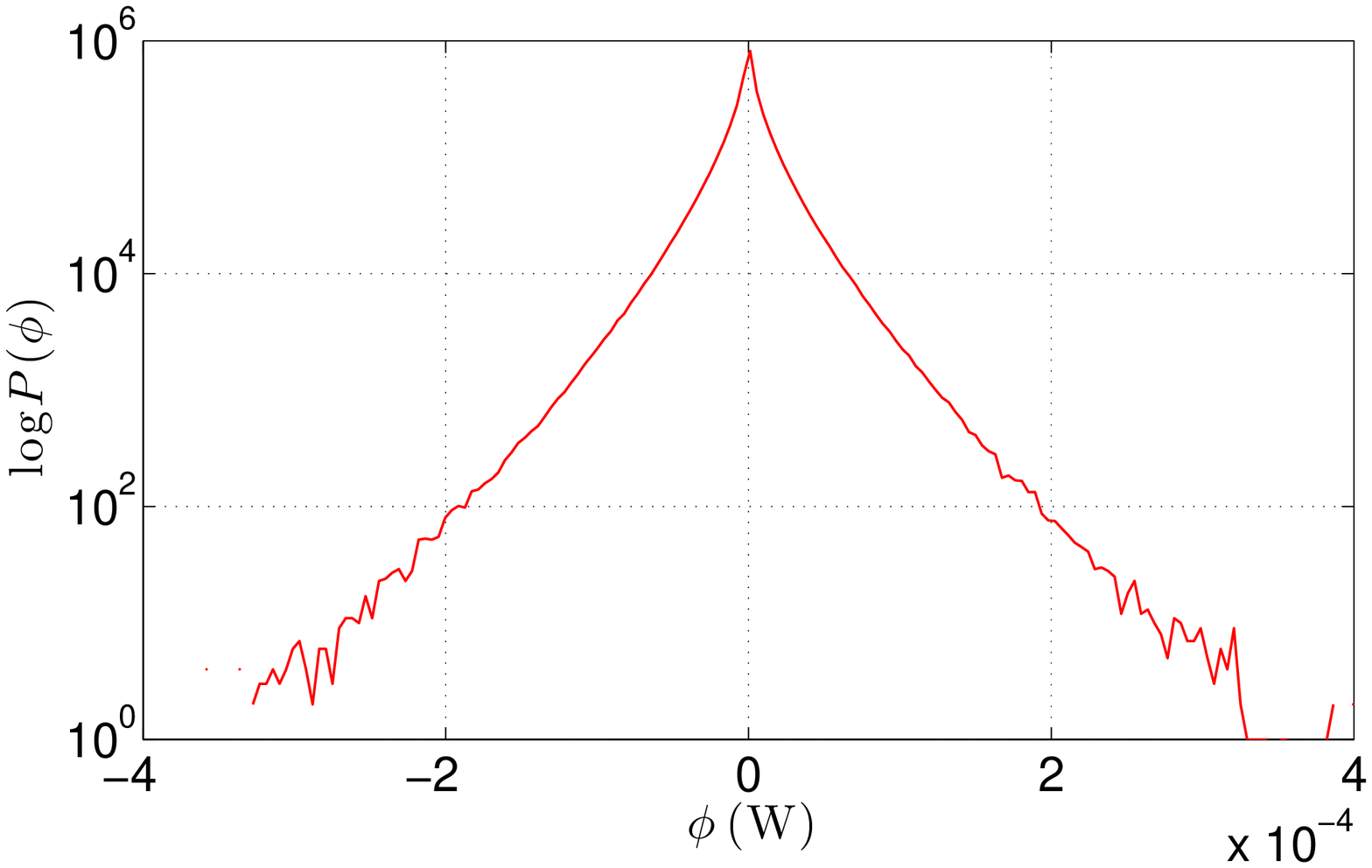}\hfill
    \includegraphics[width=7.8cm]{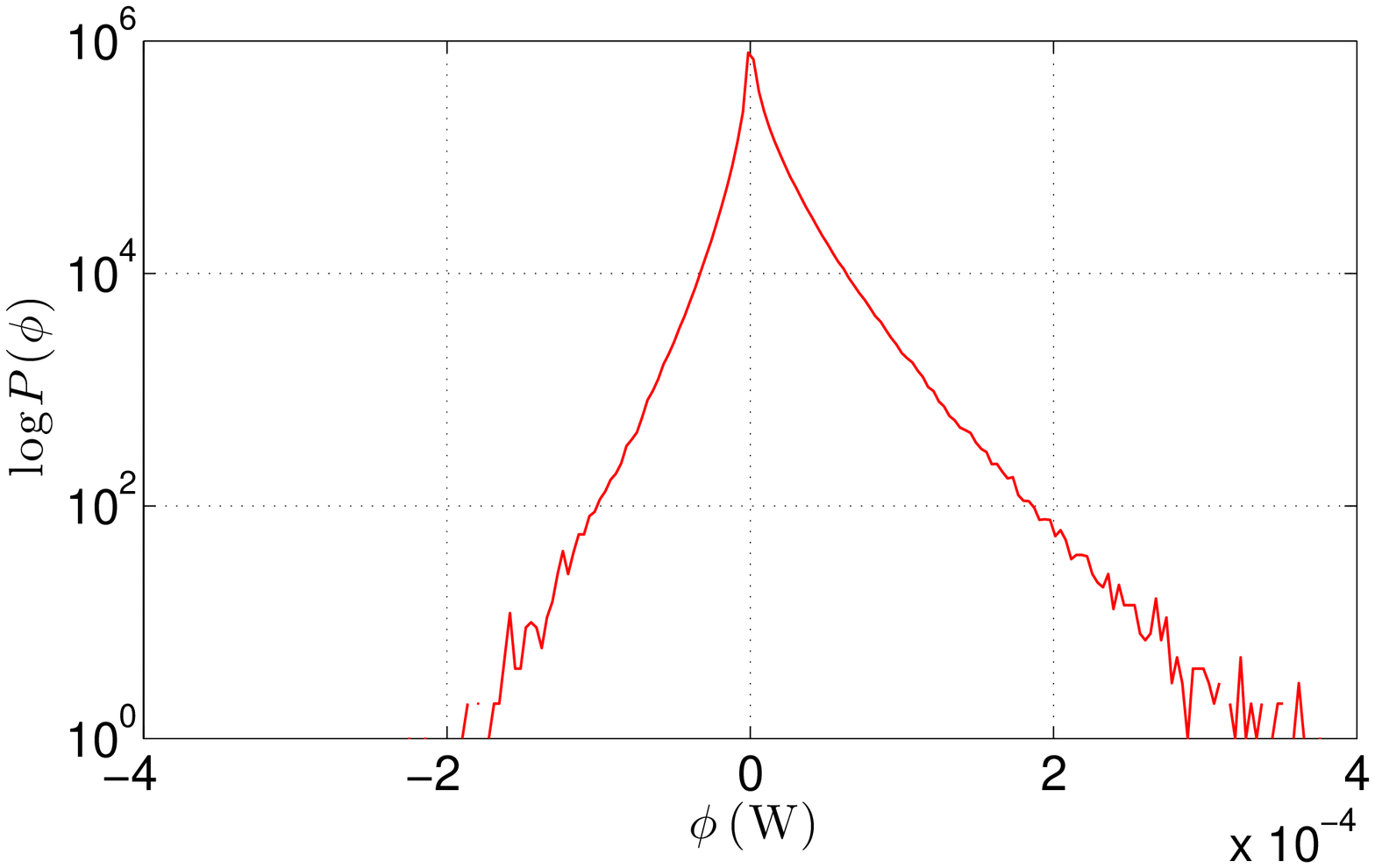}
    \caption{Histograms of the instantaneous energy flux,\red{ $\phi$,} for two different values of the temperature difference, $kT_1-kT_2$
    (Left: $kT_1-kT_2\simeq 1.7\,10^{-8}$~J, Right: $kT_1-kT_2\simeq 7.9\,10^{-7}$~J).} 
        \label{fig:histo}
\end{figure}

\begin{figure}[h!]
    \includegraphics[width=7.8cm]{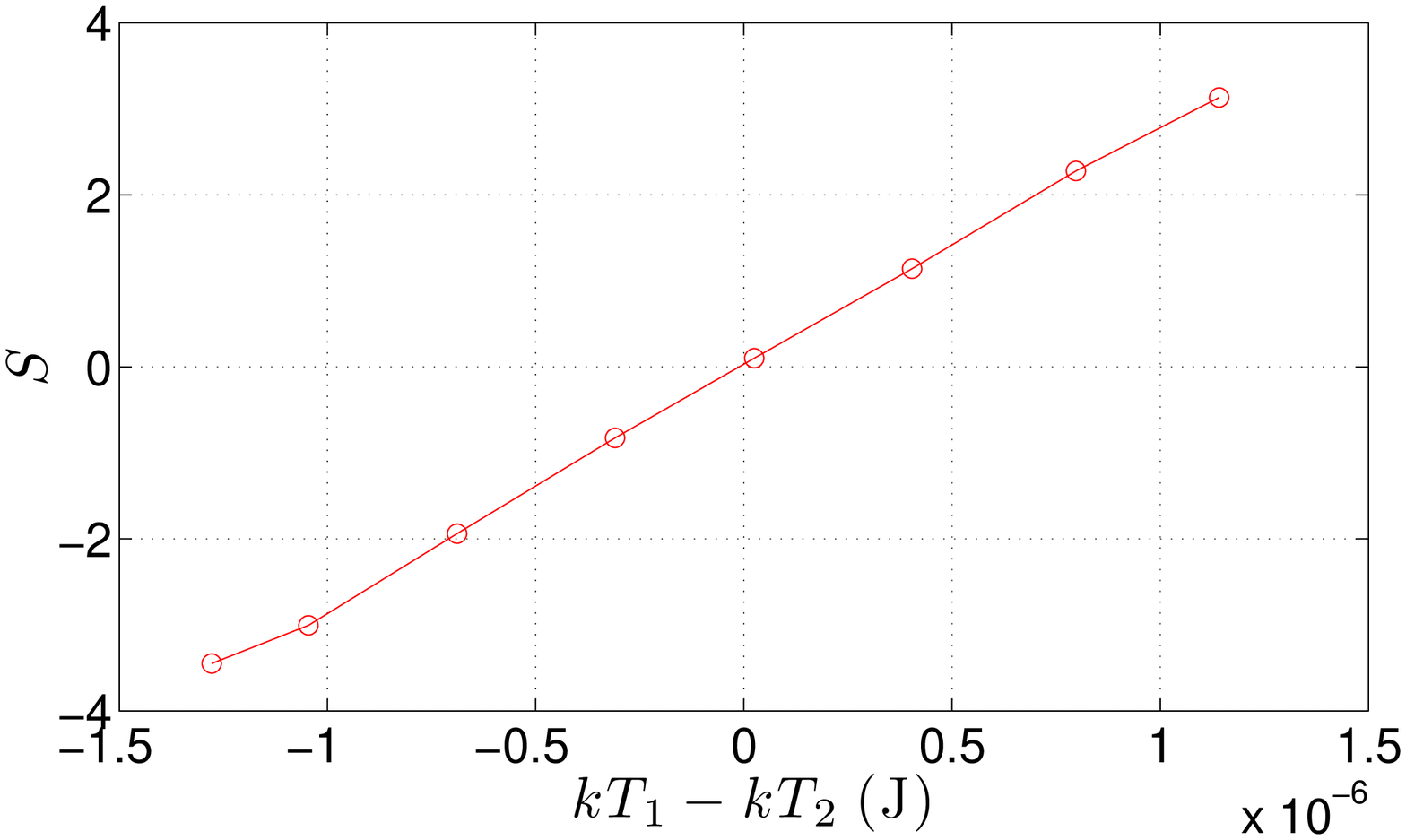}\hfill
    \includegraphics[width=7.8cm]{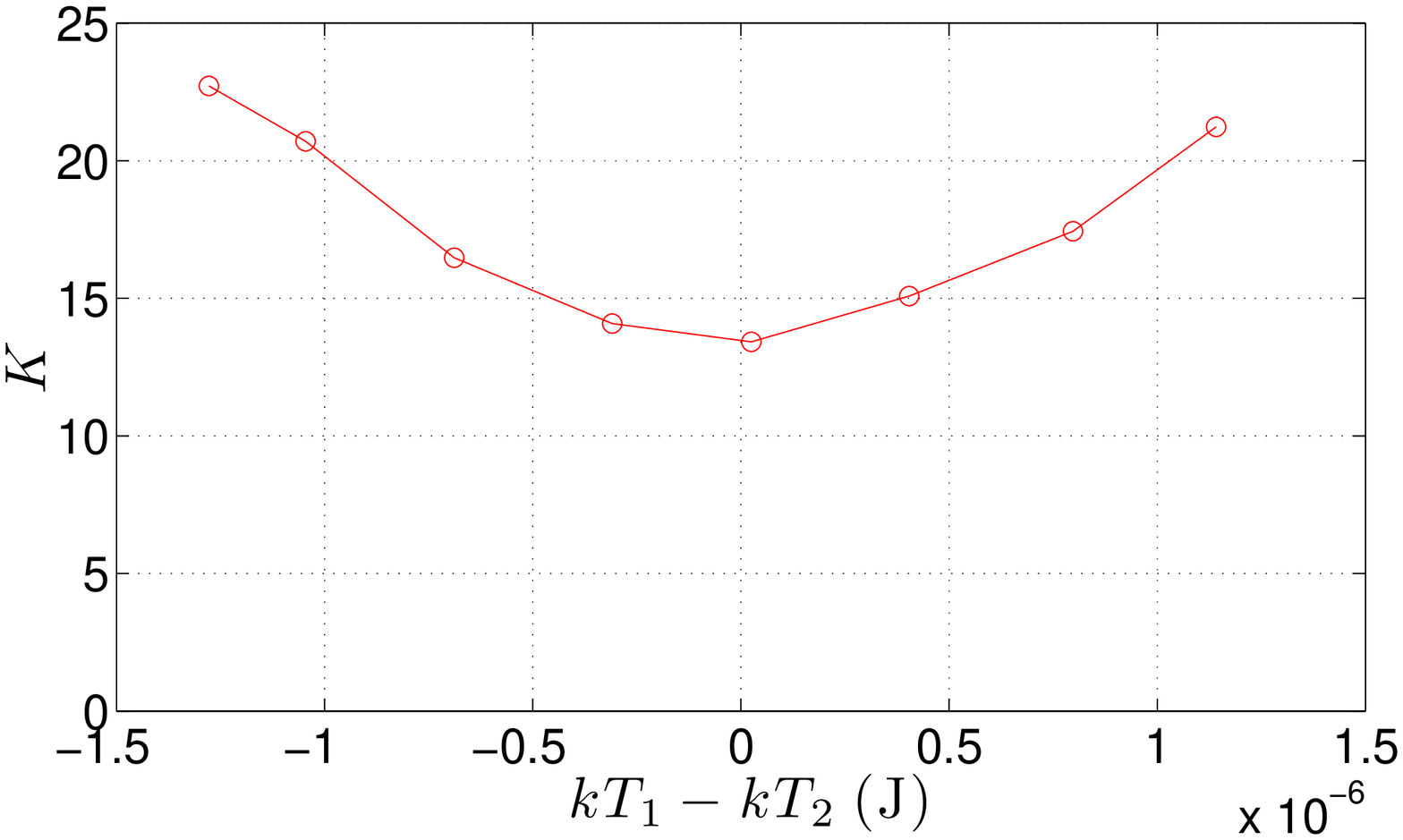}
    \caption{Skewness $S$ (left) and kurtosis $K$ (right) vs. temperature difference 
    $kT_1-kT_2$. 
        \label{fig:sk}}
\end{figure}

We report in \rep{figure }{Fig.~}\ref{fig:histo}
\red{histograms of the instantaneous energy flux, $\phi$}.
\rep{T}{We observe that t}he histograms are highly non-Gaussian and\red{,} in general\red{,} asymmetric\rem{, as shown by two examples at distinct $\Delta T$}.\rem{ \\
\noindent}
For vanishing small temperature difference, $kT_1-kT_2\simeq 0$ (Fig.~\ref{fig:histo}, left), $\overline\phi \simeq 0$ (Eq.~\ref{eq:fourier}).
We observe, in this case, that the histogram is symmetric and, accordingly, \rem{Mean flux cancels when temperature difference is zero, }that the most probable value of $\phi$ is \rem{also }zero\rem{, i.e. the histogram is symmetric (figure \ref{fig3} left)}. 
In contrast, when the temperature difference, $kT_1-kT_2$ departs significantly from 0 (Fig.~\ref{fig:histo}, right), $\overline\phi \neq 0$ (Eq.~\ref{eq:fourier}), and
the histogram is significantly asymmetric: for $kT_1 > kT_2$, the probability of a given (positive) flux $\phi$ from the {\it hot} source to the {\it cold} one
is larger than the probability of the same (negative) flux in the opposite direction. In average, the {\it heat} flows from the {\it hot} to the {\it cold} blade. 
However, \rem{A striking feature appears when temperatures in both reservoirs differ.}\red{strikingly, in spite of the asymmetry of this flow,}
\rem{The mean flux is non-zero but} the most probable \red{value of the flux $\phi$ }\rep{is always}{remains} zero\rem{: the histograms are asymmetric (figure \ref{fig3} right)}. \\
\indent To characterize these statistics at the simplest level, \red{we compute }the skewness, $S$, and kurtosis, $K$, coefficients
of \red{the distributions of }$\phi$ \rem{are computed }for different values of temperature difference $kT_1-kT_2$ \red{ (Fig.~\ref{fig:sk})}.
\rep{These are}{The coefficients, $S$ and $K$, are defined to be} the adimensional third and fourth moments \red{of the distribution}: $S={\overline{ {\Delta \phi }^3}}/{\overline{{\Delta \phi}^2}^{3/2}}$ and $K=\overline{{\Delta \phi}^4}/\overline{{\Delta \phi}^{2}}^2$,
with $\Delta \phi=\phi-\overline{\phi}$. \rem{These shape parameters are computed and plotted with respect to $kT_1-kT_2$ in figure \ref{fig:sk}.}
Note that \rep{the skewness}{$S$} and \rep{kurtosis}{K} are respectively odd and even functions \rep{versus $\Delta T$}{of $kT_1-kT_2$},
as required by \red{the }symmetry\red{ of the experimental configuration}.

\rep{At}{For} \rep{$\Delta T=0$}{$kT_1 \simeq kT_2$}, we have $S=0$ and $K$ \rep{is minimal around}{reaches a minimum of about} $14$.
Such value of \rep{the kurtosis}{$K$} is unusually \rep{high}{large} \rep{. It is respectively}{if compared to the values} $3$ and $6$ \rep{for}{associated with}
Gaussian and exponential distributions.
\section{Fluctuation Theorem}
\label{sec:fluctuation theorem}

\begin{figure}[h!]
    \includegraphics[width=7.8cm]{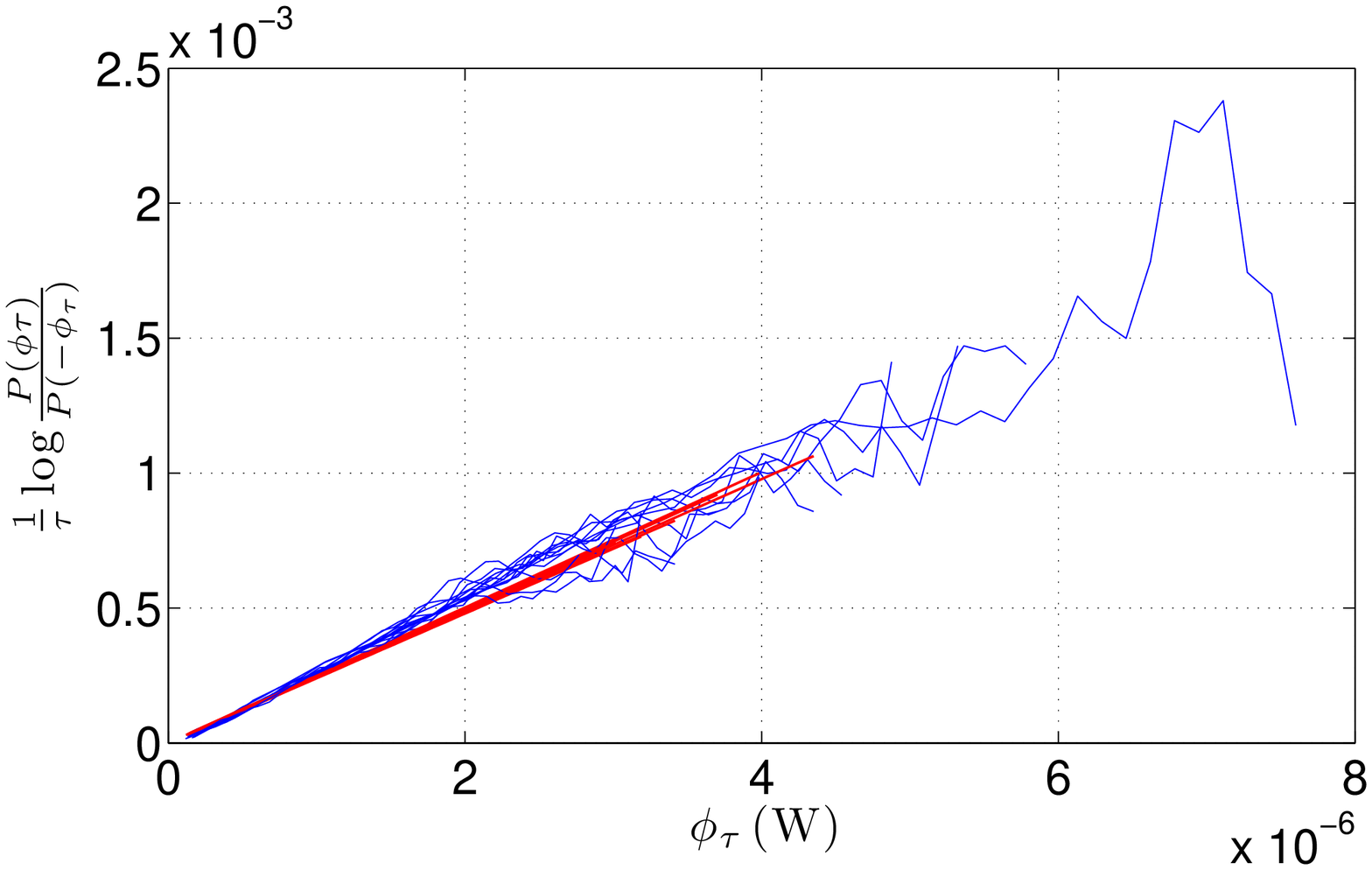}\hfill
    \includegraphics[width=7.8cm]{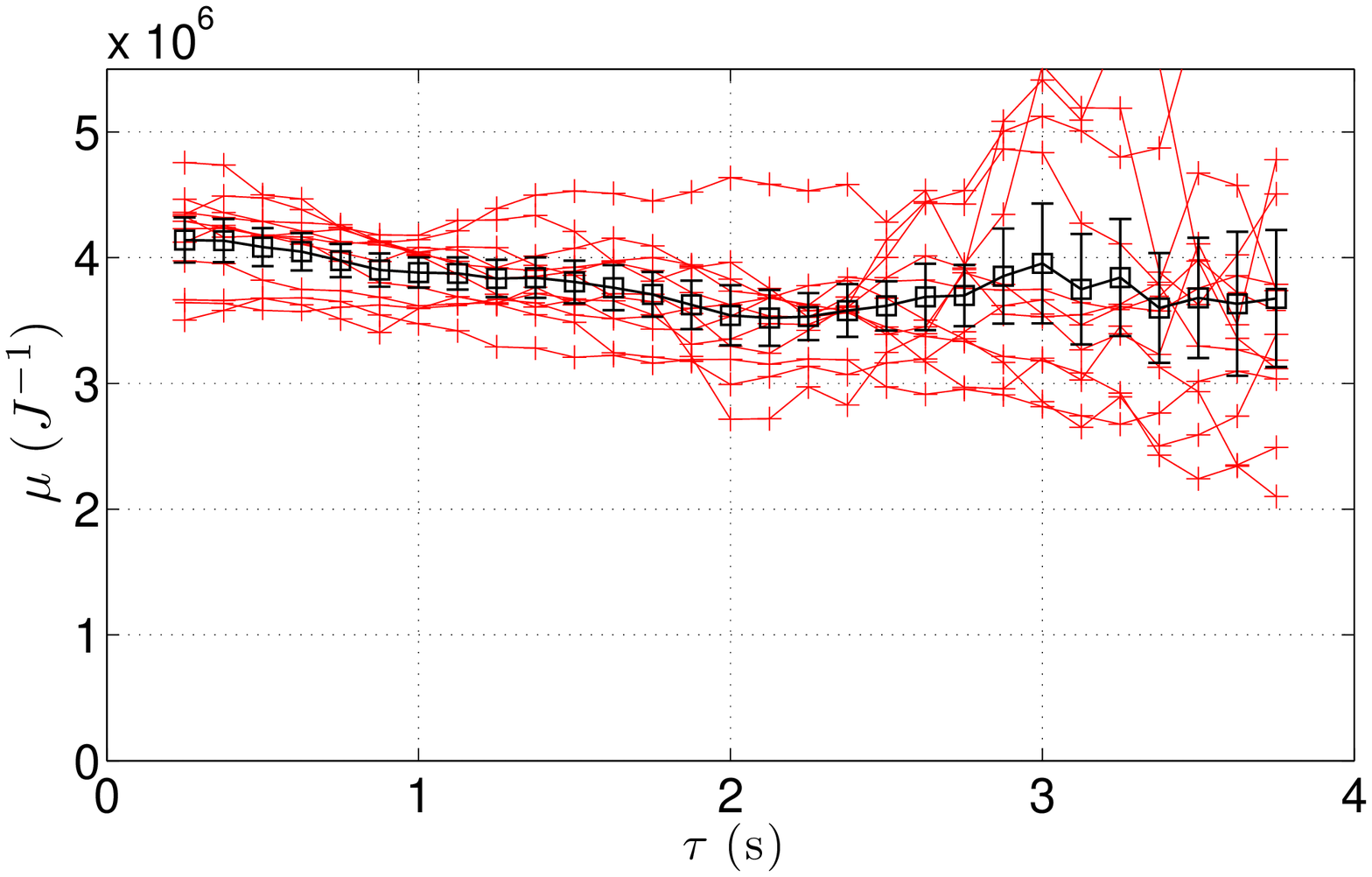}
\caption{Left: Asymmetry function vs. coarse-grained energy flux $\phi_\tau$.\newline 
Right: coefficient $\mu$ vs. $\tau$ (crosses \textcolor{red}{{\boldmath$+$}}, 1-hour time series; squares {\boldmath$\Box$}, average)
} 
\label{fig:ft}
\end{figure}

\rep{In the previous section, histograms of $\phi$ and measurements of the kurtosis and skewness parameters for various $\Delta T$ have been produced.
T}{As shown in Sec.~\ref{sec:fluctuations}, t}he distribution of $\phi$, characterized by its moments, $S$ and $K$,
happens to be\red{,} in general\red{,} asymmetric and, always, very \rep{large}{broad}.
The asymmetry of an energy flux, which reflects the asymmetry of the energy flux, is, in some situations, described by the so-called Gallavotti-Cohen Fluctuation Theorem \cite{gallavotti-cohen} (FT), that generalizes the $2^{\rm{nd}}$ principle. Typically, it is the situation where the energy flux can be related to an entropy creation rate, at a certain (fixed) temperature. Among the requirements of the FT, temperature and entropy must be well defined, which is far from obvious in a dissipative systems like granular gases. The FT compares the probability that the time coarse-grained entropy increases or decreases by the same amount during a certain (long) duration.\\ 
For the present purpose, the FT would relate, in a prospective way, the asymmetry of the distribution to the irreversibility of the dynamics and, thus, to dissipation. In our experimental configuration, dissipation mainly refers to energy losses by Joule effect in the resistors of inelastic collisions. If expressed in terms of the energy flux $\phi_{\tau}$ exchanged during a time-lag $\tau$, the FT reads in the limit of large $\tau$: 
\begin{equation}
\log\frac{P(\phi_\tau)}{P(-\phi_\tau)}={\mu \tau \phi_\tau},
\label{eq:ft}
\end{equation}
with $\phi_{\tau}=\frac{1}{\tau}\int_{\tau} \phi(t) dt=\frac{1}{\tau}\int_{\tau} (e_1+e_2)I dt$.
In this detailed form, the FT states that the relative statistical weight of positive over negative fluxes increases exponentially with $\phi_\tau$. 
In \rep{figure }{Fig.~}\ref{fig:ft} (left), the so-called asymmetry function\red{,} $\frac{1}{\tau}\log\frac{P(\phi_\tau)}{P(-\phi_\tau)}$\red{,}
is plotted against  $\phi_\tau$ for increasing values of $\tau$\rep{,}, with the best linear fit at the origin.  

\begin{figure}[h!]
\begin{center}
    \includegraphics[width=7.9cm]{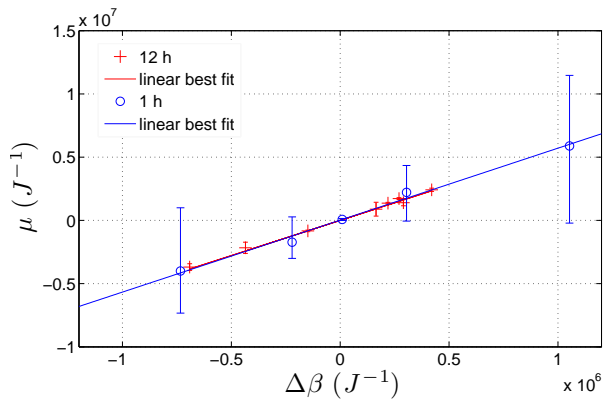}
    \end{center}
\caption{Coefficient $\mu$ vs. $\Delta \beta$.
Crosses and circles correspond to two different measurement campaigns, with $1$ or $12$ hour-long time series. }
    \label{fig:xft}
\end{figure}

The experimental data reveal a good agreement with the linear dependence of Eq.~(\ref{eq:ft}) when $\tau$ is large enough. \\
\red{For each 1-hour time series, we determine the slope and report the associated values of $\mu$ as function of $\tau$ (crosses \textcolor{red}{{\boldmath$+$}}, Fig.~\ref{fig:ft}, right. Note that, in order to limit the influence of statistical noise, }we consider the slope at the origin).
Finally, we average 12 curves of $\mu(\tau)$, calculated over 12 consecutive 1-hour \rep{files}{time series} and estimate the error bars from the standard deviation (squares {\boldmath$\Box$}, Fig.~\ref{fig:ft}, right). We observe that, the coefficient $\mu$, determined as explained above, reaches an asymptotic value plateau for large $\tau$, before the increase of the statistical noise. The transient at small $\tau$ is not investigated, only the asymptotic value at large $\tau$ is relevant, from the viewpoint of the FT. This asymptotic value is called $\mu$ from now on. 
It can be seen in figure \ref{fig:ft} (right) that the typical time of convergence is several seconds, whereas the correlation time of the exchange with the gas is a few dozen of ms. This is a consequence of the intermittency of $\phi$ (revealed by the large value of the kurtosis).
The slow convergence of any statistical quantities makes the measurement of $\mu$ especially difficult.
\\
We point out that the strict linearity of the asymmetry function with the existence of an asymptote $\mu$ at large $\tau$ implies the validity of the FT for the energy flux. A note of caution has to de added though. We cannot exclude that a larger sample would reveal a third order term in Eq.~\ref{eq:ft}, and therefore discard the FT. This restriction, always present in experimental work on the subject, must be mentioned (see a detailed discussion on this point in \cite{aumaitre2001}). However, within our statistical resolution, we exhibit a strong indication that the FT holds but not a proof in the strict sense.
\\
It is now interesting to assess the dependence of the asymptotic value of $\mu$ on the temperature difference. 
In the absence of losses, during a quasi-static transport process, between two large equilibrium thermostats, the relation expected would be: $\mu = \Delta\,\beta$, with $\Delta \beta = \frac{1}{k\,T_1}-\frac{1}{k\,T_2}$. This is not without reminding a version of the Fluctuation Theorem derived for heat flux by Jarzynski et al. named {\it exchange fluctuation theorem} (XFT) \cite{jarzynski2004}, where $\mu = \Delta\,\beta$. 
We observe in Fig.~\ref{fig:xft} that our experimental value of $\mu$ indeed is proportional to $\Delta \beta$ defined, we remind, from the fluctuations of the angular velocities of the blades. We notice, however, that slope of the best linear fit is experimentally $5.69$ and not 1. 

Our observation also differs on that point from the recent measurements on energy transport between Nyquist resistors at equilibrium at various temperatures coupled by a capacitor, i.e. non-dissipative \cite{ciliberto2013}. These last measurements are consistent with the XFT.\\
At this point, the question of this discrepancy remains open. Among the necessary conditions listed above for the FT with the slope $\mu = \Delta\,\beta$, a certain number are more than doubtful. The relatively low dimensionality of the heat reservoirs, 
or simply the non-equilibrium nature of these reservoirs are all causes that certainly can be invoked. In any case, the XFT need not to apply here.
\section{Discussion and conclusion}
\label{sec:discussion}
We achieved an experimental situation which makes possible to study the stationary transport of energy
between two granular heat reservoirs.
The experimental set-up presented here is technically very simple, thanks to small electromechanical devices.
The crafty use of DC-motors, simultaneously as actuator and probe, avoids calibration difficulties. The two sub-systems, that mimic independent dissipative steady-state thermostats, are designed to be identical. We note that, besides the mean energy flux, all quantities relative to the fluctuations are symmetric for temperature difference inversion (Fig. \ref{fig:sk} and \ref{fig:xft}). This is a strong indication that the twin system is symmetric.\\
Let us first comment that we do not assess the properties of the gases directly, but rather the dynamical properties
of the probes in contact with the granular system. Thus, as is, $\phi$ is the instantaneous energy flux between the two
probes and not between the gases. These two instantaneous energy fluxes differ from one another by the changes in the kinetic energy
of the blades. 
However, in temporal average, in the stationary regime, the kinetic energy of the blades remains constant
and $\overline\phi$ is indeed the energy flux between the two reservoirs. 
\\
We reported that $\overline{\phi}$ is proportional to the temperature difference between the two probes. This means, as the $kT$ measure is proportional to the temperature of the granular gas, that the mean energy flux between the two granular gas baths is linear. 
This feature is similar to what would be observed for equilibrium systems.
\\
Interestingly, in our dissipative system, the fluctuations exhibit surprising features: they are very intermittent,
in the sense that intense events are highly represented in the statistics (as shown by the large value of the kurtosis, $K \geq 14$);
the most probable value of $\phi$ remains $0$ even when, in average, $\overline{\phi} \ne 0$ (this is shown by the fact that the skewness
departs from $0$ when the temperature difference is increased).
When comparing these two studies, one must keep in mind that the conditions are distinct in several ways. In particular, our thermostats are dissipative and far from the thermodynamic limit (small number of grains, relatively large mean free path). It is difficult at this point to give the origin of the peculiarities observed here. It would certainly be interesting to build an experiment allowing to vary either the number of degrees of freedom alone, or the dissipation alone.
\\
Focusing further on the fluctuations of the flux, we observe that the asymmetry functions are linear. 
In other words, the probability of observing a coarse grained flux in one direction over the probability of the opposite during the same time grows exponentially with the amplitude of this flux. In the limit of large time-lags, the slope converges to an asymptotic value, which is already remarkable. Although not a proof in the strict sense, it is a strong indication for the Fluctuation Theorem. We emphasise on the fact that the context of our experiment does not fulfill the hypothesis needed for the FT to apply.  \\
Beyond, the asymptotic slope is measured for various temperature gradients. It is found proportional to the inverse temperature difference, but not equal. The value of the slope is still a question to be addressed in the future, experimentally or numerically. The departure on this point from a theory like the XFT, or the measurements of Ciliberto et al. can be various, as the context is different in several ways: the thermostats are out of equilibrium, implying that temperature is an effective temperature; the number of degrees of freedom of the thermostats is relatively small; the coupling is dissipative. \\
These results really demand a better understanding of the conditions in which the FT could be applied in the context of dissipative stationary processes. Macroscopic experiments such as the present one definitely helps, as they possibly allow to vary parameters. \\
As we pointed out, the validity of the FT in the present context is far from obvious despite the results shown in this article, and the subject is far from being exhausted. The system presented deserves to be improved. Using seemingly simple experimental technics, it gives access to fundamental questions on non-equilibrium statistical mechanics which deserve careful examination and interpretation. We believe it allows for a critical study of the influence of the different hypothesis of the FT out of its validity framework.
\ack
We gratefully acknowledge S.~Ciliberto, S.~Auma\^itre, J.-Y.~Chastaing, J.-P.~Zaygel, A.~B\'erut, J. Farago, and K.~Mallick for discussions and advises, as well as numerous students and researchers of the ENS-Lyon. We highly appreciate I.~Fr\'erot, M.~Clusel, and J.-C.~G\'eminard  for pertinent questioning, and rereading of the manuscript.

\end{document}